\documentclass[acmsmall, authorversion]{acmart}

\AtBeginDocument{%
  \providecommand\BibTeX{{%
    \normalfont B\kern-0.5em{\scshape i\kern-0.25em b}\kern-0.8em\TeX}}}

\setcopyright{acmlicensed}
\acmJournal{PACMHCI}
\acmYear{2022} \acmVolume{6} \acmNumber{CHI PLAY} \acmArticle{246} \acmMonth{10} \acmPrice{15.00}\acmDOI{10.1145/3549509}

\usepackage{enumitem}
\newcommand{\compresslist}{
  \setlength{\itemsep}{1pt}
  \setlength{\parskip}{0pt}
  \setlength{\parsep}{0pt}
}
\usepackage{tabularx}
\usepackage{mathtools}
\usepackage{balance}
\newcommand{\comm}[1]{}

\begin{document}

\title{Outpace Reality: A Novel Augmented-Walking Technique for Virtual Reality Games}

\author{Sebastian Cmentowski}
\email{sebastian.cmentowski@uni-due.de}
\orcid{0000-0003-4555-6187}
\affiliation{%
  \department{High-Performance Computing}
  \institution{University of Duisburg-Essen}
  \city{Duisburg}
  \country{Germany}
}

\author{Fabian Kievelitz}
\email{fabian.kievelitz@stud.uni-due.de}
\orcid{0000-0002-4369-081X}
\affiliation{%
  \department{High-Performance Computing}
  \institution{University of Duisburg-Essen}
  \city{Duisburg}
  \country{Germany}
}

\author{Jens Kr\"uger}
\email{jens.krueger@uni-due.de}
\orcid{0000-0002-9197-0613}
\affiliation{%
\department{High-Performance Computing}
  \institution{University of Duisburg-Essen}
  \city{Duisburg}
  \country{Germany}
}

\begin{abstract}
The size of most virtual environments exceeds the tracking space available for physical walking. One solution to this disparity is to extend the available walking range by augmenting users' actual movements. However, the resulting increase in visual flow can easily cause cybersickness. Therefore, we present a novel augmented-walking approach for virtual reality games. Our core concept is a virtual tunnel that spans the entire travel distance when viewed from the outside. However, its interior is only a fraction as long, allowing users to cover the distance by real walking. Whereas the tunnel hides the visual flow from the applied movement acceleration, windows on the tunnel's walls still reveal the actual expedited motion. Our evaluation reveals that our approach avoids cybersickness while enhancing physical activity and preserving presence. We finish our paper with a discussion of the design considerations and limitations of our proposed locomotion technique.
\end{abstract}



\begin{teaserfigure}
  \includegraphics[width=\textwidth]{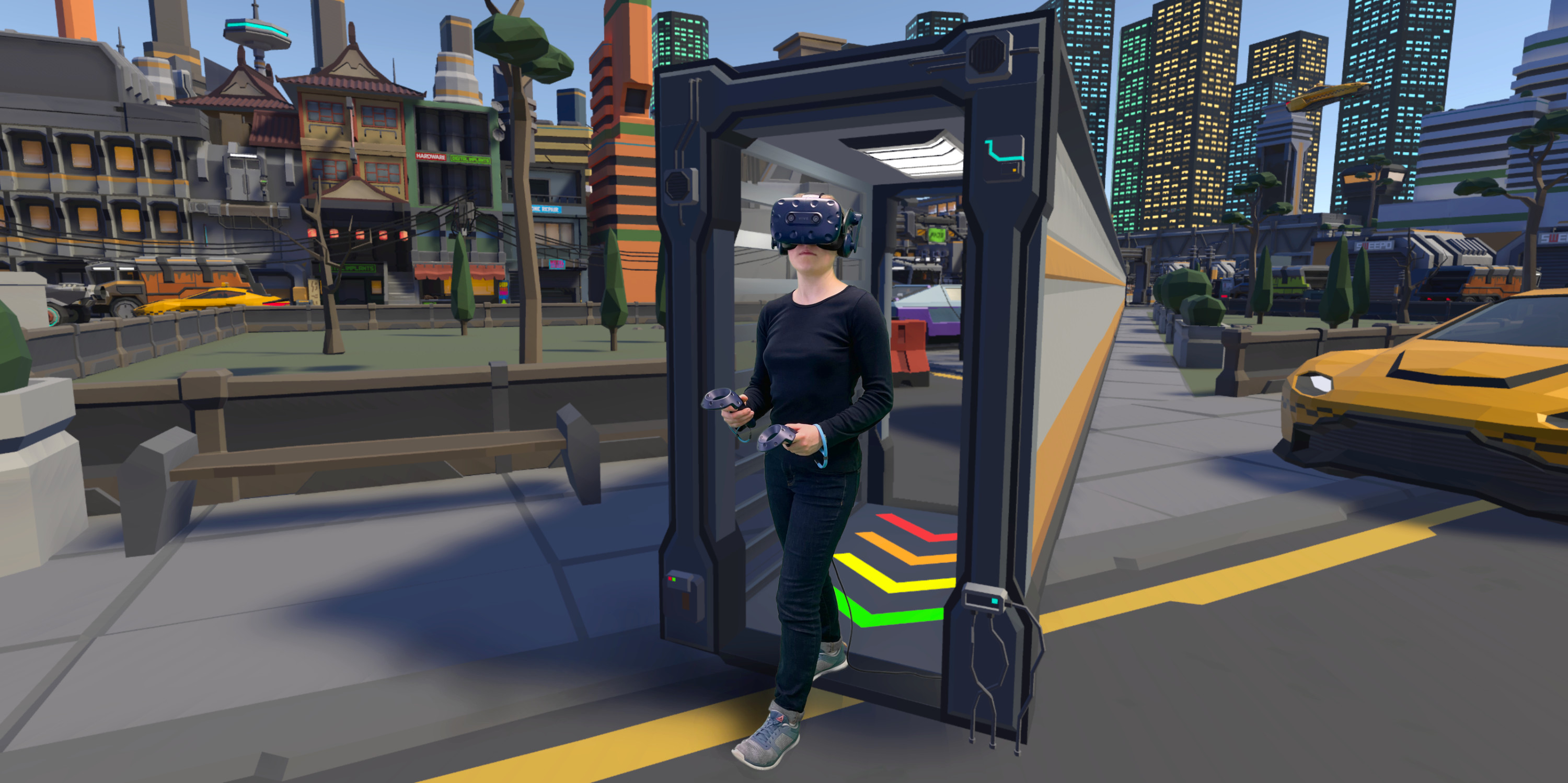}
  \caption{Our proposed navigation technique accelerates users' physical steps to traverse larger distances. The novel tunnel concept prevents cybersickness by shielding users from excessive visual flow. Windows in the tunnel's walls provide a direct view of the augmented movement.}
  \Description{Female participant (visually merged into the virtual world) leaves our presented virtual tunnel. The background is a futuristic city with modern buildings and flying cars.}
  \label{fig:teaser}
\end{teaserfigure}

\maketitle

\section{Introduction}
Walking is of central importance in our daily lives. We visit friends, go to work, or hike in the nearby forest --- for most of our activities, we need to travel to a different location. Also, walking is one of the most effective ways to keep us healthy and active. All this sums up to an average walking distance between four and eight kilometers every day~\cite{bassett2010_pedometer}. The same also applies to virtual reality, which reflects or even extends the real world. Whether it is an engaging game, geographical visualization, or educational lesson, most immersive experiences require the users to travel between distant points of interest. Despite decades of research on virtual locomotion techniques, real walking is still considered the gold standard. It not only feels most natural to users but also offers a range of valuable benefits, including an increased physical activity~\cite{cmentowski2021_sneaking}, the avoidance of cybersickness~\cite{ruddle2009_benefits}, and a superior spatial orientation~\cite{ruddle2011_walking}.

The available tracking space of room-scale VR systems is usually highly limited. Living room setups, in particular, rarely exceed a few square meters in size. This technical constraint severely limits the usefulness of real walking and increases the risk of collisions with physical obstacles. Therefore, past work presented different approaches to increase the available walking range, such as augmenting users' steps~\cite{interrante2007_boots, williams2006updating}. Increasing the movements in the virtual world allows users to travel larger scenarios fast and effectively~\cite{williams2006updating}. However, the necessary transitional gains also introduce a deviation between real and virtual locomotion, which is a typical source of cybersickness. Also, this approach would augment otherwise unperceived motions, such as head bobbing or tracking errors. Therefore, the \textit{Seven League Boots} concept by Interrante et al.~\cite{interrante2007_boots} limits the transitional gain to the users' forward movement alone. Whereas this improvement effectively reduces cybersickness through head bobbing, previous studies highlighted several other remaining issues that diminish the practical applicability of this technique. 

Firstly, the necessary detection of the users' intended movement direction often remains challenging in practice~\cite{abtahi2019giant, williams2008design}. Also, a clear distinction between augmented traveling and local navigation is crucial, as scaling stationary head movements is known to cause disorientation and cybersickness~\cite{williams2019scaling}. Furthermore, translational gains were repeatedly shown to reduce navigational accuracy and increase the necessary workload~\cite{abtahi2019giant, cirio2009magic, nabioyuni2015evaluation, wilson2018object}. Finally, cybersickness remains a severe issue. Even after eliminating any unwanted sideways movements, accelerated physical walking still increases the perceived visual flow in the entire field of view. Consequently, past research demonstrated a direct correlation between the applied gain factor and the severeness of cybersickness symptoms~\cite{tirado2019_gains}. In sum, augmented walking in the current form was rarely tested with gain factors exceeding $10x$. Instead, Abtahi et al.~\cite{abtahi2019giant} suggest limiting translational gains to $3x$ at most. This constraint severely limits practical applicability for larger scenarios, particularly in combination with typically restricted play areas.

Our research addresses the issues above by introducing a novel accelerated-walking approach for virtual scenarios. The proposed navigation concept prevents cybersickness by vastly reducing the perceived visual flow from the augmented forward movement. Also, our technique avoids the loss in accuracy by restricting the accelerated travel to a fixed path. Finally, our design makes it easy to distinguish clearly between local walking and long-distance navigation. The core element of our prototype is a virtual tunnel providing an easy and short way of traveling along a direct route between the users' current position and a predetermined target location (see Figure~\ref{fig:teaser}). This tunnel appears to span the entire length to the navigation goal when viewed from the outside. In contrast, the tunnel's interior is just a fraction as long to enable the users to traverse it easily by real walking. For instance, a tunnel with a length of $75m$ shrinks to $2.5m$ physical walking distance when applying a gain factor of $30x$. We achieve this impression by combining multiple concepts, such as portals for the tunnel's entry and exit.

As the users walk through the tunnel, their forward movement is scaled so that they reach their target position when leaving the compressed inner tunnel. Nevertheless, this augmentation stays primarily unnoticed as the players gain the impression of just walking physically through the short tunnel. This concept allows for expanding the users' movements drastically without causing cybersickness through an increased visual flow. However, a perfect shortened tunnel illusion would lead to similar disorientation and lack of impression of the traveled distance known from portals and other relocations. Therefore, our virtual tunnel features window slits that provide a direct peripheral view of the actual expedited movement. Thus, the users gain an impression of moving faster than usual through the windows while the rest of the tunnel serves as a visual rest frame preventing cybersickness effects. The windows' optimal shape and size were fine-tuned in a participatory design phase to account for individual preferences and perceptual differences.

Further, we validate our presented navigation technique against the widely established point \& click teleport. Being a completely virtual travel concept featuring instant and free locomotion within the boundaries of the VR game, teleportation can be considered the direct opposite of our approach. Twenty-five participants used both techniques in an immersive game featuring distant points of interest and local tasks. The results of this within-subject study reveal that our presented technique increases physical activity while preventing cybersickness and preserving high levels of presence. Additionally, the experiments show that our concept is easier to learn and use than the teleport technique. In the last part of our paper, we discuss the resulting design considerations and limitations of our proposed locomotion technique.

In sum, the main contributions of our work are the following:
\begin{enumerate}[leftmargin=*]\compresslist
    \item a novel navigation approach for straight paths based on augmented real walking
    \item a within-subject study to compare the performance and reception of our locomotion technique against the teleport
    \item a set of design considerations covering the strengths and limitations of our approach.
\end{enumerate}

\section{Related Work}

As our work belongs to VR games and locomotion research, we first introduce basic concepts such as immersion, presence, and cybersickness. Next, we cover the current state of research on VR locomotion in general. Finally, we close this section with a deeper look at prior work on augmented movements and portal-based navigation.

VR systems allow users to dive into a virtual, nonexistent world. Thereby, the VR setup's technical ability to suppress the physical surroundings is called \textit{immersion}~\cite{cairns2014immersion, biocca1995_immersion, sherman2003_understanding}. If done correctly, the users of such highly immersive systems can experience a feeling of being present in the virtual world~\cite{heeter1992_being}. This sensation of \textit{presence} is often a decisive factor in assessing the quality of a VR experience~\cite{slater2003_note, ijsselsteijn2000_presence, lombard_2000_presence}. The work by Slater et al.~\cite{slater1995_steps} focuses explicitly on how VR locomotion impacts the perceived presence.

VR applications often suffer from inducing symptoms of cybersickness~\cite{laviola2000_cybersickness, hettinger1992_visually}. Whereas past research has shown that immersive experiences can still be enjoyable despite the occurrence of cybersickness~\cite{vonmammen2016_cybersick}, most developers avoid causing discomfort to the users. In the literature, cybersickness is often used interchangeably with the effect of simulator sickness~\cite{kolasinski1995_simulator}. However, both are different subsets of motion sickness~\cite{money1970_sickness, ohyama2007_autonomic}. Simulator sickness~\cite{stanney1997_cybersickness} is typically caused by technical problems of a misconfigured simulator and leads to mild oculomotor and nausea symptoms~\cite{kennedy1989_simulator}.

In contrast, the sources of cybersickness lie within a sensory mismatch of the human vestibular and visual systems~\cite{reason1975_motion}. The exact reasons remain unclear and are manifested in three theories: sensory conflict theory, poison theory, and postural instability theory~\cite{laviola2000_cybersickness}. Cybersickness leads to rather severe symptoms~\cite{reason1975_motion}, such as disorientation and nausea~\cite{stanney1997_cybersickness}. It is caused by various factors, including technical issues and configurational problems, caused by wrong eye distance or vergence. Also, fast-moving objects combined with a large field of view (FOV) increase the perceived visual flow and contribute to the formation of symptoms~\cite{lee2017_flow}. Consequently, limiting the FOV can help in avoiding cybersickness~\cite{fernandes2016_fov, lin2002_fov}. Finally, Hettinger et al.~\cite{hettinger1990_vection} proposed vection as another causative factor. Vection is a phenomenon where the visual system induces a feeling of self-movement despite the lack of other motion cues. A typical example is watching a nearby train accelerating through the windows of a standing train. In sum, it appears that discrepancies between real movement and observed virtual motion, paired with a high optical flow rate, amplify potential cybersickness. 

\subsection{Locomotion}

Moving through virtual environments is essential for most VR scenarios:  exploring extensive worlds, searching for distant points of interest, or traveling to spatially separated locations~\cite{mcmahan2014_interaction, tan2001_flying}. Numerous locomotion techniques have been proposed throughout decades of VR research, each with its strengths and weaknesses. As the holy grail of a universal navigation concept may not even exist, locomotion research remains vital to provide fitting approaches for every use case. Considering the vast quantity of existing locomotion techniques, we limit this section to a brief overview. For a detailed summary, we point to existing taxonomies and reviews~\cite{boletsis2017_typology, boletsis2019_comparison, laviola2017_interfaces, alzayer2020_survey}, such as the recent collection \textit{LocomotionVault}~\cite{diluca2021_locomotionvault}. 

Despite years of research, natural walking is still considered the best locomotion approach~\cite{alzayer2020_survey, ruddle2009_benefits}. Its intuitive and presence-preserving navigation provide unmatched benefits. Especially when the locomotion task matches the real-world counterpart, real walking is more efficient than virtual travel~\cite{suma2009_evaluation}. Additionally, physical walking helps users construct a cognitive map of the environment~\cite{ruddle2011_walking}. Finally, using the users' real movements for virtual navigation also enable other novel concepts such as using the gait as an input modality~\cite{cmentowski2021_sneaking, hoppe2019_vrsneaky}. However, the dimensions of the real play area usually limit room-scale tracking to a few square meters.

Therefore, purely virtual navigation techniques decouple the real and virtual movements to achieve an unlimited range of travel. Nevertheless, approaches involving continuous virtual locomotion without a physical counterpart tend to induce cybersickness~\cite{habgood2017_lessons, smolyanskiy2017_view}. A solution is using short and fast movements without acceleration, as these are usually unproblematic~\cite{medeiros2016_speed, yao2014_oculus}. The most prominent approach is the instant teleport~\cite{bozgeyikli2016_teleport}, which is superior to gamepad locomotion~\cite{frommel2017_locomotion}. However, teleportation might break the presence~\cite{bowman1997_travel} and cause spatial disorientation, as the instant relocations limit the users' ability to estimate the traveled distance~\cite{bakker2003_spatial, bowman1997_travel}. Another popular concept is the world-in-miniature~\cite{stoakley1995_wim}, which has been subject to many research efforts. For a review of the extensive design space, we point to the work by Danyluk et al.~\cite{danyluk2021_wim}. Despite numerous proposed concepts, virtual locomotion often fails to achieve the same perceptual qualities as natural walking~\cite{usoh1999_walking}. Therefore, other approaches aim to preserve the benefits of real walking while overcoming the physical playspace's restrictions. These approaches fall into three categories:

The first group encompasses hardware solutions, such as omnidirectional treadmills~\cite{darken1997_treadmill, vijayakar2002_treadport}. However, these often bulky devices have rarely found their way into broader adoption. Next are concepts that mimic walking through physical mockups, such as walking-in-place~\cite{slater1995_steps, tregillus2016_vrstep}. As initial approaches did not feel as natural as real walking~\cite{usoh1999_walking}, later research~\cite{feasel2008_llcmwip, templeman1999_wip, yan2004_walking} improved the underlying algorithms, e.g., by involving gait-related biomechanics~\cite{wendt2010_gud} and switching seamlessly between real walking and walking-in-place~\cite{bhandari2017_legomotion}.

Finally, the last subset comprises hybrid solutions that are based on real walking but extend the available range of movement. Our proposed navigation concept belongs to this category. The most renowned representatives of this subset are redirection techniques, such as redirected walking~\cite{razzaque2001_redirected, razzaque2005_redirected}. They work by unconsciously deviating the virtual locomotion from the real movements. For example, applying slight virtual rotations provokes the users to move in circles while gaining the feeling of walking a straight line. Throughout the last few years, numerous improvements~\cite{engel2008_controller, grechkin2016_detection, langbehn2016_reorientation} have been proposed, including refining the detection thresholds for different users and conditions~\cite{williams2019_gain} or introducing alignment-based redirections that reduce collisions with the physical environment~\cite{williams2021_arc}. For an in-depth review of the state of research, we point to the work by Nilsson et al.~\cite{nilsson2018_redirected} and Suma et al.~\cite{suma2012_taxonomy}. Finally, the \textit{Space Bender} concept by Simeone et al.~\cite{simeone2020_bender} achieves a comparable effect by overtly altering the environment instead of the users' movements. Nabiyouni and Bowman~\cite{nabiyouni2016_taxonomy}, as well as Cardoso and Perrotta~\cite{cardoso2019survey}, provide more detailed analyses of these and other walking-based locomotion techniques that are not relevant to our work.

\subsection{Portals in Virtual Environments}

A limited amount of prior research utilized portals for different purposes, such as folding the virtual world~\cite{cheng2015_turkdeck}, without explicitly focussing on their use as a navigation concept. The \textit{Worlds-in-Wedges} concept~\cite{nam2019_wedges} used volumetric portals to split the surrounding environment, allowing users to view several worlds simultaneously. Finally, portals were also used to transfer between different sceneries~\cite{husung2019_portals} or from an entry environment to the target scenario~\cite{steinicke2009_transitional, steinicke2010_transitions}.

In locomotion research, portals were mainly employed to redirect users back to the playspace's center and extend the available walking range. In this context, Misztal et al.~\cite{misztal2020_portals} used portals to prevent cable twists by applying a physical 180° rotation without altering the virtual position or orientation. Freitag et al.~\cite{freitag2014_portals} also forced users to turn 180°, whereas Liu et al.~\cite{liu2018_redirected} used a combination of smaller pre- and postportal rotational gains to achieve similar effects. Both works combined the rotational aspect with a positional change to steer the users away from the playspace's boundaries and reduce the frequency of necessary resets. Finally, the \textit{Arch-Explore}~\cite{bruder2009_arch} technique combined portal navigation with the world-in-miniature concept: users choose a location on a miniature model and then use a static portal to visit the selected site.

Despite the promising applications, these works also reveal the drawbacks of using portals for VR locomotion. Freitag et al.~\cite{freitag2014_portals} reported that the portal condition performed worst concerning spatial orientation compared to instant teleport and flying. This loss of orientation is also a key gameplay feature behind the game Portal~\cite{portal}. A possible explanation for the poor comprehensibility of portals might reside in the human ability to form a cognitive map of the direct surroundings. Suma et al.~\cite{suma2012_impossible} used \textit{impossible spaces}, i.e., virtual scenarios with overlapping geometry, to demonstrate the flexibility of human spatial perception. Most of the time, the participants did not recognize geometrical inconsistencies. The authors speculated that the human mind overlooks most impossibilities as long as the local view stays consistent. This finding might explain the difficulties with portals such as these destroy the users' locally consistent view.

\subsection{Augmenting physical movements}

Prior work extended the available walking range by amplifying the mapping between the users' real and virtual movements. Applying such translational gains~\cite{williams2006updating}, which are mostly used for various redirected walking concepts, remains unnoticed below a detection threshold around $1.25x$~\cite{grechkin2016_detection, steinicke2009estimation}. Within this scope, previous works examined the impact on gait parameters~\cite{janeh2017analyses} and object selection performance~\cite{wilson2018object}. As larger gain factors also amplify secondary motions, such as head bobbing, the \textit{Seven League Boots} concept by Interrante et al.~\cite{interrante2007_boots} limited scaling to the intended forward movement during walking. Similarly, Xie et al.~\cite{xie2010_gain} also used translational gains combined with resetting. Another approach by Bolte et al.~\cite{bolte2011_jumper} augments the forward movement of a detected jump to travel further.

However, applying translational gains introduces deviations between the virtual and the real movement, which leads to cybersickness~\cite{christou2017_steering}. Rietzler et al.~\cite{rietzler2020_telewalk} combined translational and rotational gains for a perceivable redirection approach but reported significantly more cybersickness symptoms than with teleportation. Similar effects on sickness levels were also reported for large translational gains, where almost $50\%$ of subjects reported considerable symptoms~\cite{tirado2019_gains}. Apart from the experienced cybersickness, prior research also emphasized adverse effects on accuracy and the need for a clear distinction of local and accelerated walking~\cite{williams2019scaling}. Finally, Abtahi et al.~\cite{abtahi2019giant} also suggested an increased acclimation time, as users may only experience the applied speed gain after beginning walking. Instead of putting up with these drawbacks of translational gains, one might scale the complete player by increasing the height and eye distance accordingly~\cite{cmentowski2019_outstanding, krekhov2018_gullivr}. This concept preserves the proportions between real and virtual movements. Instead of causing cybersickness, it leads to the impression of walking through a toy world and allows for comfortable accelerated locomotion.

\section{Tunnel-based Locomotion Concept}

Our proposed navigation technique extends the walking range without causing cybersickness by combining two concepts: We apply a translational gain in the target direction and use a virtual portal to create a space-bending illusion. Similar to the \textit{Seven League Boots} technique~\cite{interrante2007_boots}, our concept increases the users' movement in the forward direction. However, we do not infer the intended walking direction but only augment the movement in the direct line between start and target. Consequently, our navigation technique focuses on fixed and straight routes between a predetermined start point $p_s$ and an endpoint $p_e$. This design decision prevents the loss in accuracy that is typically observed with scaled movements. As users travel between the start and endpoint, they cover the virtual distance $d(p_s, p_e)$. However, in reality, augmenting the movement speed by a transitional gain factor $a_g$ reduces the walking distance to $d(p_s, p_e)/a_g$. For example, traveling $60m$ from one point of interest to another requires $2m$ in the actual play space with a gain factor of $30x$.

\begin{figure*}[t!]
\centering
\includegraphics[width=\columnwidth]{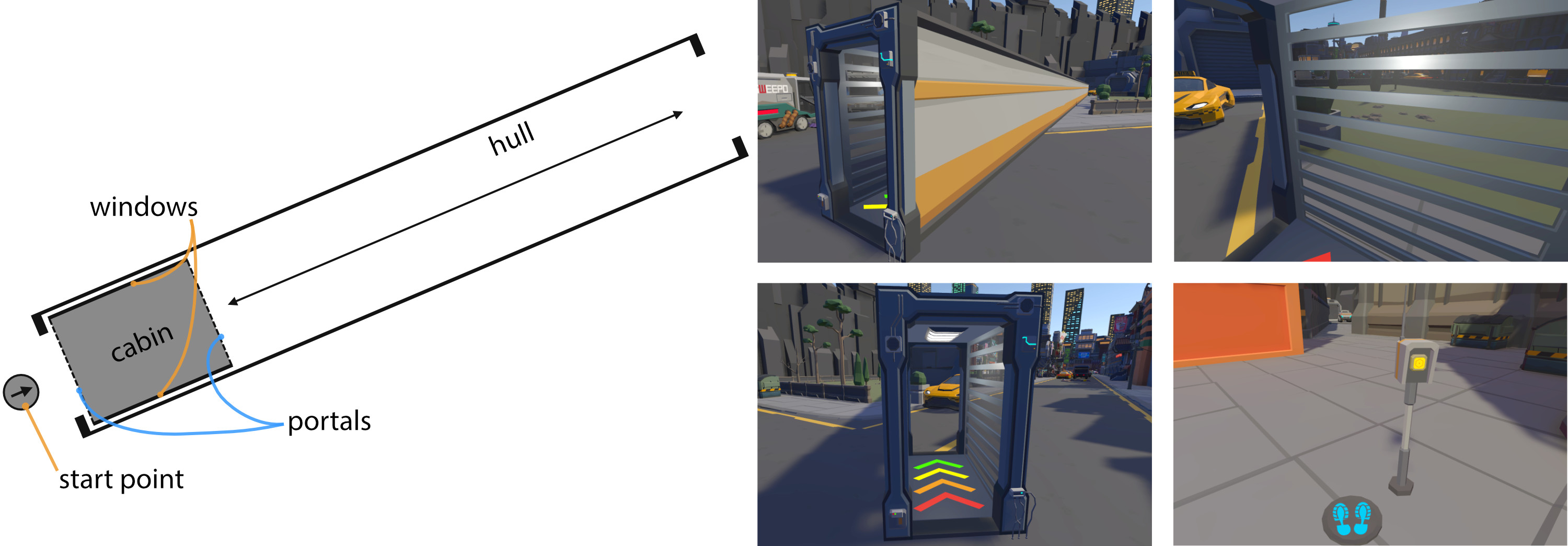}
\caption{The virtual tunnel, as the core of our novel locomotion technique: The exterior \textit{hull} spans from the start point to the target. While users walk through the \textit{cabin}, it is moved along the tunnel to transport the users to their destination. The cabin's walls and portals create an optical illusion, whereas the windows provide a view of the accelerated movement. Users initiate the tunnel by standing on a platform and pressing a button.}
\label{fig:tunnel}
 \Description{On the left is a schematic of our tunnel concept featuring the corresponding labels. On the right are four images illustrating the hull, cabin, windows, and starting point.}
\end{figure*}

Previous research showed that although this kind of translational movement effectively enhances the movement range of real walking, it may also cause cybersickness. Therefore, we constructed a virtual tunnel to increase the available walking range through translational gains while avoiding adverse side effects and preserving the advantages of natural walking to physical activity and spatial orientation (see Figure~\ref{fig:tunnel}). This virtual tunnel consists of an exterior component, the \textit{hull}, and an interior element, the \textit{cabin}. The tunnel's hull is mainly used to illustrate the actual travel distance before the users enter the tunnel. The hull features an entry and exit arch and exterior tunnel walls and roof that span the entire distance $d(p_s, p_e)$. Therefore, the users may investigate the tunnel from the outside and understand that it spans from their position $p_s$ to their target $p_e$. The tunnel's cabin is an enclosed room fitting into the tunnel's hull. The cabin's length is $d(p_s, p_e)/a_g$, as this is the actual physical distance for the users to walk. Whereas the walls to both sides, the floor, and the roof consist of solid surfaces, the two sides covering the cabin's entrance and exit are portals displaying the actual tunnel's ends. This construction provides the impression of a shortened passage within the actual long tunnel. To the players, the two parts hull and cabin merge into one tunnel, which appears to be very long on the outside, but relatively short when looking through it.

When the users enter the tunnel, the rig, determining their position in the world, is made a child object of the cabin. As the users walk through the passage, the cabin is moved accordingly with an increased speed of $v * a_g$. As a result, the cabin reaches the tunnel's far end just as the players arrive at the cabin's exit portal. This concept is similar to an elevator moving within its shaft. As the portal conforms with the actual view, the users do not notice a visual difference when leaving the cabin. This construction provides the perfect impression of walking through a short tunnel and exiting at the rear side, arriving at the distant target. The players are shielded from the increased optical flow arising from the augmented movement, so we do not expect any cybersickness.

However, the tunnel hides the actual long-distance travel in this form and would probably diminish the players' spatial orientation and feeling of traveled distance. Therefore, we deliberately break this perfect illusion by adding windows to the cabin's sides. These expose the actual virtual locomotion speed. By doing so, we aim to provide a better impression of the travel experience. Still, we assume that the rest of the cabin, i.e., the portals and the walls, suffice to provide a steady reference point reflecting the actual walking speed and preventing cybersickness. Related concepts in prior research are dynamic FOV limiters that shield users from increased visual flow~\cite{fernandes2016_fov, lin2002_fov}. However, such techniques overtly reduce the players' view of the virtual environment and could diminish the player experience. In contrast, our approach uses diegetic and thematically embedded geometry to achieve a similar effect while preserving a partial view of the occluded areas through the windows. Thereby, we aim to reduce any potential disturbance to a minimum.

Of course, determining the best shape and size of the windows is essential. Both extreme cases do not provide the wanted effects: tiny windows fail to raise an impression of fast movements, whereas huge windows, covering the complete walls, increase the visual flow and potentially induce cybersickness. Therefore, we evaluated different window sizes and shapes in a participatory design phase with four participants with a mean age of $40$ ($SD=17.62$). Despite the limited subject count due to COVID-19 restrictions, we made sure to include different degrees of VR expertise and susceptibility to cybersickness. All tested designs featured the same solid wall material and an almost transparent glass surface serving as a window to the outside environment. Although traditional rectangular cutouts provided the best impression of the travel experience, extended use sometimes induced mild symptoms of vection-based cybersickness. We assume that the missing visual reference frame when the windows fill the users' view might have caused this experience. For this reason, a design consisting of horizontal window stripes was favored by most participants. Additionally, the resulting effective window size is larger than traditional windows but split into small stripes, providing an optimal steady reference.

Finally, we aimed to increase the impression of traveling a longer distance by adding decorative and animated elements. Therefore, we used arrows with switching colors on the cabin's floor to provide a similar effect as the acceleration fields in many games, such a Mario Kart~\cite{marioKart}. Also, we added animations when the tunnel appears and disappears. After enabling the tunnel, the users see it rising to half its height from the floor. Then, the tunnel expands in length until it reaches the target destination before rising to full height. Finally, the tunnel's doors are opened and allow the users to traverse it. After the users arrive at their destination, the doors close automatically, and the tunnel is retracted into the ground. Whereas our concept would still work without these effects, we presume that they further strengthen the players' impression of the actual travel distance. Also, the brief pause induced by the animations helps distinguish the local navigation and the long-distance travel.

In sum, our virtual tunnel is a novel locomotion technique focussing on straight and predetermined paths. It aims to prevent the adverse side effects of large translational gains by hiding most of the visual flow and providing a steady reference frame while providing a restricted impression of the movement through fixed windows.

\section{Evaluation}

We conducted a lab study to evaluate our proposed navigation concept. Therefore, we used an urban virtual environment to compare the augmented-walking approach against the commonly used \textit{point and teleport} technique~\cite{bozgeyikli2016_teleport}. We were interested in the qualitative and quantitative differences in the efficacy of both methods, especially regarding usability, spatial orientation, and cybersickness.

\subsection{Research Questions and Hypotheses}

Our primary research goal is to explore the differences between our introduced augmented-walking technique and instant teleportation. Apart from being probably the most commonly used locomotion technique in recent games and applications, we mainly chose the teleport for its completely contrastive properties. Whereas our approach limits navigation between points of interest to a fixed and straight route, teleportation enables players to navigate the world freely within the boundaries of the reachable game world. This traveling happens instantaneously compared to our concept, which is based on real walking and involves short animations. Finally, the teleport is well-known for the shallow learning curve and the absence of cybersickness. Altogether, the different characteristics promise intriguing insights into the individual strengths and weaknesses. Therefore, we compare both approaches regarding the eight performance metrics proposed by Bowman et al.~\cite{bowman1998_methodology} to assess the efficacy of locomotion techniques: speed, accuracy, spatial awareness, ease of learning, ease of use, information gathering, presence, and user comfort.

\subsubsection{User Comfort}

The primary motivation of our work is to avoid the possible cybersickness typically observed with disparities between real and virtual velocity. Our tunnel shields the user from most of the increased visual flow, so we assume that it effectively prevents cybersickness.

\subsubsection{Speed and Accuracy}

In contrast to the user's comfort, a high travel speed is not a crucial requirement for VR locomotion techniques. Instead, it should suit the intended use case and allow users to move through the virtual environment without causing boredom or fatigue. As our navigation concept is based on real walking, we expect to observe significantly longer travel times compared to instant teleportation. At the same time, we suppose that users walk further and feel more physically active. 

Due to the conceptual differences between both locomotion techniques, we do not compare the navigational accuracy. In contrast to the teleport, users must choose their final destination for the virtual tunnel directly at the start. Precisely setting a navigation target at a distance of more than 50 meters is not doable with the visual pointer used for teleportation. A possible solution would be to use a dedicated interface, such as a minimap or WIM, to determine the target. However, this design decision introduces an additional interaction concept, which might bias the evaluation. Therefore, we exclude this step from our study and instead employ tunnels with fixed start and end positions. The users initiate the tunnel by standing on a small platform and pressing a knob in the virtual scenario, similar to an elevator button. This design decision has implications for the user experience, as discussed later in the paper.

\subsubsection{Spatial Awareness and Information Gathering}

Real walking positively impacts the human ability to generate a cognitive map of the surroundings and thus leads to improved spatial awareness and a better impression of the environment. Whereas these findings suggest a positive effect of our augmented-walking technique, the surrounding tunnel hides part of the scenery during the travel and might reduce the benefit. Still, we assume that the windows provide a good view of the scene and significantly benefit spatial orientation compared to immediate teleportations.

\subsubsection{Ease of Learning and Ease of Use}

Apart from providing an effective, comprehensive, and comfortable travel experience, locomotion techniques must also be easy to learn and use. Our navigation approach merely requires the user to open the tunnel and walk through it. Both are activities we perform every day, e.g., when using an elevator. Therefore, we assume that the tunnel is reasonably easy to master. However, the teleport is also known for simplicity and is the preferred choice in action-intensive VR games. Thus, it remains an open research question how these two techniques compare regarding usability and mastery.

\subsubsection{Presence}

Finally, the success of immersive VR experiences largely depends on the users' perceived feeling of presence. This sensation is influenced by most aspects of the VR application, including the utilized locomotion technique. Even though real walking is commonly believed to be the most natural movement concept, past research has revealed mixed results concerning the effects on perceived presence. Whereas some walking-based approaches elicited higher presence levels~\cite{krekhov2018_gullivr, cmentowski2019_outstanding, usoh1999_walking}, other comparing studies could not confirm such effects~\cite{langbehn2018evaluation}. The lacking consensus prevents us from hypothesizing potential significant effects. Instead, we pose a research question to investigate the influence of the two conditions on presence.

\begin{figure*}[t!]
\centering
\includegraphics[width=\columnwidth]{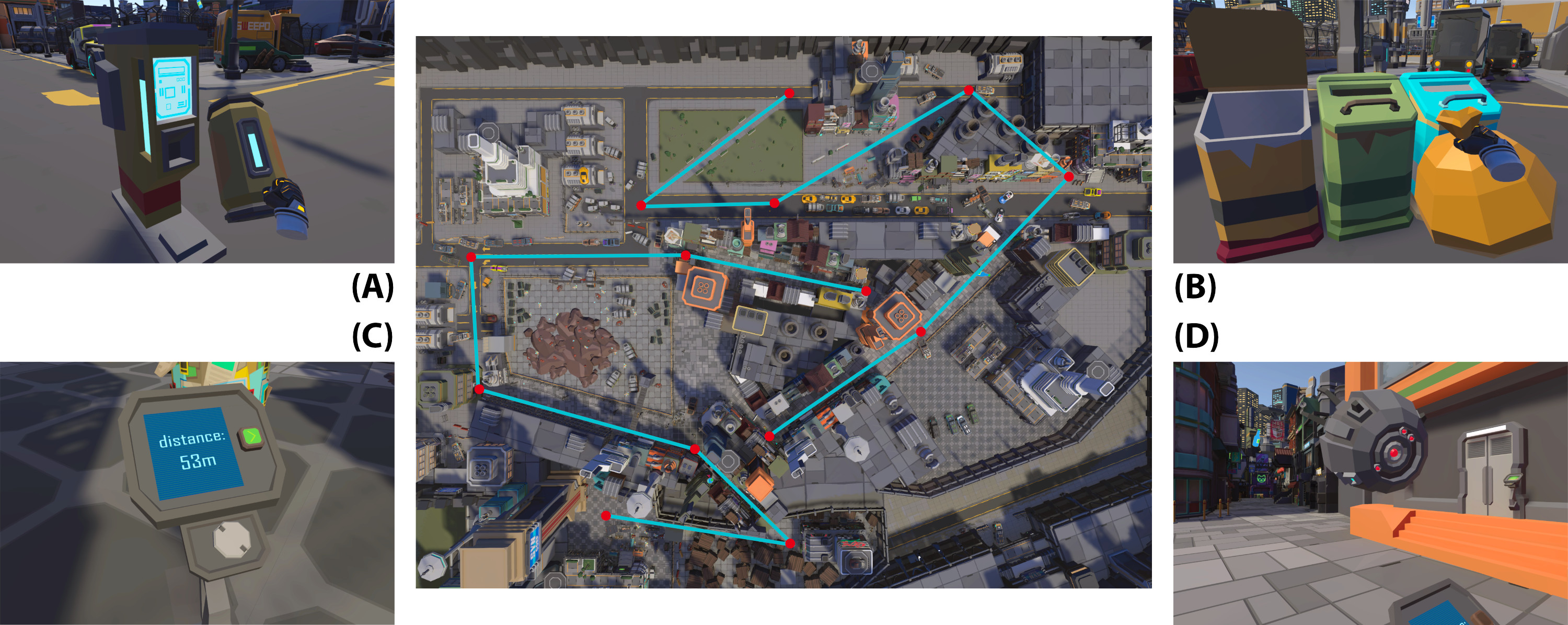}
\caption{Overview of the testbed environment, including the two levels $L1$ and $L2$, the \textit{energy supply} task (A), the \textit{garbage collection} task (B), the distance estimation (C), and the instruction drone (D).}
\label{fig:map}
  \Description{In the center is a top-view map of our testbed scenario (with both routes marked). Around the map, there are four images illustrating the various scenario elements.}
\end{figure*}

To summarize, our hypotheses and research questions are:

\noindent\begin{itemize}[leftmargin=*]\compresslist
    \item H1: The presented augmented-walking approach does not induce cybersickness.
    \item H2: Compared to teleportation, our locomotion technique significantly increases the walked distance and the overall travel time.
    \item H3: The introduced tunnel concept significantly benefits spatial orientation and gives a better impression of the traveled route compared to teleportation.
    \item RQ1: How does our locomotion technique compare to teleportation regarding mastery and ease of control?
    \item RQ2: Do both navigation approaches differ concerning perceived presence?
\end{itemize}

\subsection{Testbed Scenario}

We realized our testbed game to compare both navigation techniques using the Unity game engine 2021.2.0b4~\cite{unity}. Our virtual environment is a futuristic city featuring narrow alleys, wider streets, and ample open spaces (see Figure~\ref{fig:map}). The style is similar to dystopian cyberpunk scenarios to allow for easy integration of the required mechanisms. The subjects can move only within the designated play area throughout the study, consisting of seven points of interest, connected by six direct paths. The scene's geometry, i.e., buildings, vehicles, fences, or barriers, is designed to form a single path without requiring explicit boundaries. 

As we aimed for a within-subject design, we designed two nearly identical levels $L1$ and $L2$. Both encompass different streets in the same testbed scenario and share most properties, such as the distance and angle between the points of interest. We exchanged only the visuals, such as the surrounding buildings, to avoid repetition effects. Additionally, we created two similar local tasks for the points of interest. Both rely on the same principle: sorting colored items into matching slots. In the first task, \textit{energy supply}, the users take three energy cells from a central point of issuance and place them into the correct loading facilities. In the second task, \textit{garbage collection}, the users collect three garbage bags and place them into one of the three garbage bins. No additional locomotion technique is required for the tunnel condition, as designing the environment with the lab dimensions in mind ensures that all interactables are reachable by real walking within the boundaries of the play area. Our task design combines the investigated long-range navigation with local interactions and thus resembles typical game plots. Also, it ensures good comparability as both tasks use the same principle and employ a similar walking pattern.

Each level is structured identically: The subjects begin the experience at the starting location of the respective level $L1$ or $L2$. A nonhumanoid drone provides the necessary instructions for completing the assigned task. Depending on the level, the users must complete either the energy supply or garbage collection task. Then, they are introduced to the locomotion technique, i.e., augmented walking or teleport, and use it to travel to the second point of interest. After arriving there, the subjects estimate their traveled distance. The resulting estimation errors are later analyzed for differences between the conditions. Thereby, the six paths between the seven checkpoints have three different lengths repeating twice: $60m$, $75m$, $45m$, $75m$, $45m$, and $60m$. As we applied a constant gain factor of $30x$, this design leads to tunnel lengths between $1.5m$ and $2.5m$. After entering their distance guess into a terminal, the users continue the pattern --- task, travel, distance estimation --- until the last task at the final checkpoint is completed. Finally, the drone reappears and debriefs the subjects.

\subsection{Procedures and Applied Measures}

We conducted a within-subject study and applied a cross-over design to avoid sequence effects, i.e., subjects either started with teleport or augmented walking. Additionally, we wanted to avoid repetition effects through learning the local task. Therefore, we counterbalanced the order of tasks. That said, each subject started with a combination of one task and one locomotion technique and replayed the opposite combination in the second round. The study took 45 minutes on average and was conducted in our $16 m^2$ lab using an HTC Vive Pro~\cite{vive}.

In the beginning, we informed the subjects about the overall procedure and the general study's goals. Before the first round, we assessed the subjects' general information, such as gender, age, and prior VR and gaming experience. Next, we introduced the subjets to the VR setup, assisted them in adjusting the headset properly, and explained the functionality of collision warnings. After these preliminary steps, we started the first level, where the subjects received their instructions. Throughout the playthrough, we logged the relevant statistics needed for our hypotheses H2 and H3: walking distance, travel duration, and the subjects' personal distance estimations. For the travel duration, we measured the time from completing all tasks at one checkpoint until reaching the next checkpoint. Therefore, these timings also include the necessary time to invoke the tunnel. After completing the level, the subjects completed a set of questionnaires assessing their personal experience.

First, we administered the Simulator Sickness Questionnaire (SSQ)~\cite{Kennedy.1993} to test our hypothesis H1 and examine possible cybersickness effects. The SSQ consists of the three subscales nausea, oculomotor, and disorientation, using a range from 0 (none) to 3 (severe) for the individual elements. As we were interested in the influence of the locomotion technique on the perceived presence (see RQ2), we used the Presence Questionnaire (PQ) (initially developed by Witmer and Singer~\cite{Witmer.1998} and later revised by the UQO Cyberpsychology Lab~\cite{uqo2004_presence}). It allows for a closer look at the locomotion-related influences on presence and includes five subscales: realism, possibility to act, quality of interface, possibility to examine, and self-evaluation of performance (coded 0 - 6). For RQ1, exploring the usability of the navigation concepts, we assessed three subdimensions of the player experience inventory (PXI)~\cite{abeele2020development}: ease of control, autonomy, and master (all coded -3 - 3). Finally, the questionnaires were complemented by eight custom questions (coded 0 - 6). These covered spatial orientation, a key concern in H3, and other locomotion-related questions, expanding our understanding of the different locomotion techniques. Semistructured interviews finished the study to allow all subjects to share their experiences.

\setlength{\tabcolsep}{2pt}
\begin{table*}[t]
  \caption{Mean scores, standard deviations, and paired sample t-test values of the Presence Questionnaire (PQ), the Player Experience Inventory (PXI), and the Simulator Sickness Questionnaire (SSQ).}
  \label{tab:results}
  \begin{tabularx}{\textwidth}
  {>{\raggedright\arraybackslash}p{3.6cm}
  >{\centering\arraybackslash}X
  >{\centering\arraybackslash}X
  >{\centering\arraybackslash}p{1.3cm}
  >{\centering\arraybackslash}p{0.8cm}
  >{\raggedright\arraybackslash}p{0.3cm}
  >{\centering\arraybackslash}p{0.9cm}
  >{\centering\arraybackslash}p{2.3cm}
  }
    \toprule
     & Tunnel & Teleport & \\
     & M(SD) & M(SD) &$t(24)$ & $p$ & & $d$ & CI\\
    \midrule
    PQ (scale: 0 - 6)\\
    \ \ \ \ \ \ Realism & 4.18 (0.89) & 3.76 (0.92) & 2.526 & .019 & * & .505 & $[.076, .758]$\\
    \ \ \ \ \ \ Possibility to Act & 4.27 (0.91) & 4.33 (0.93) & -.417 & .680 & & -.083 & $[-.357, .237]$\\
    \ \ \ \ \ \ Interface Quality & 4.97 (0.76) & 4.93 (0.80) & .323 & .749 & & .065 & $[-.215, .295]$\\
    \ \ \ \ \ \ Possibility to Examine & 4.48 (0.98) & 4.31 (0.85) & 1.192 & .245 & & .238 & $[-.127, .474]$\\
    \ \ \ \ \ \ Performance & 4.94 (0.79) & 4.60 (0.91) & 1.846 & .077 & & .369 & $[-.040, .720]$\\
    \ \ \ \ \ \ Total & 4.45 (0.71) & 4.23 (0.71) & 2.241 & .035 & * & .448 & $[.017, .404]$\\
    PXI (scale: -3 -- 3)\\
    \ \ \ \ \ \ Mastery & 1.56 (1.04) & 1.52 (0.98) & .285 & .778 & & .057 & $[-.250, .330]$\\
    \ \ \ \ \ \ Autonomy & -0.33 (1.71) & -0.60 (1.36) & .900 & .377 & & .180 & $[-.345, .878]$\\
    \ \ \ \ \ \ Ease of Control & 2.37 (0.75) & 2.09 (0.78) & 2.929 & .007 & ** & .586 & $[.083, .477]$\\
   SSQ (scale: 0 - 3)\\
    \ \ \ \ \ \ Nausea & 13.74 (15.35) & 12.59 (17.57) & .768 & .450 & & .154 &  $[-1.931, 4.220]$\\
    \ \ \ \ \ \ Oculomotor & 17.28 (17.16) & 16.98 (16.70) & .157 & .877 & & .031 &  $[-3.683, 4.290]$\\
    \ \ \ \ \ \ Disorientation & 25.06 (38.75) & 19.49 (31.64) & 1.477 & .153 & & .295 &  $[-2.212, 13.348]$\\
    \ \ \ \ \ \ Total & 20.64 (23.06) & 18.55 (21.31) & 1.028 & .314 & & .206 &  $[-2.112, 6.301]$\\
    \bottomrule
     &&&&& \multicolumn{3}{r}{*\textit{p} $<.05$, ** \textit{p} $<.01$}\\
\end{tabularx}
\end{table*}

\section{Results}

Based on the circumstances during data collection, i.e., the COVID-19 pandemic, prior experiences from other studies, and a preliminary power analysis, we aimed for a sample size of N=25. As the significance threshold, we chose $\alpha=.05$. These factors enable the detection of effect strengths of $d=0.6$ with $80\%$ power. For our study, we applied strict hygiene measures to ensure the safety of our participants. These measures included the mandatory use of face masks, regular airings, and disinfection of the headset after every use. The 25 recruited participants (10 female, 15 male) with a mean age of 27.84 (SD=13.92) reported playing digital games at least a few times per month ($80\%$) and mainly had used VR systems before ($92\%$). Only a minority stated they were using VR on a regular basis ($24\%$).

Our study comprised two locomotion techniques, which are subsequently referred to as tunnel and teleport. Also, we used a pair of similar local tasks in two levels with equal geometrical layouts. Thus, we split the subjects randomly into four groups. Each started one locomotion technique and one task and played the opposite combination in the second round. This cross-over design accounts for potential sequence and learning effects. To confirm comparability between both local tasks, we grouped trials sharing the same locomotion technique and compared the time subjects took to complete the local assignments at every checkpoint. This average task completion time differs significantly for neither the teleport condition ($task_{energy}=72.09$ (SD=18.65); $task_{garbage}=65.06$ (SD=19.74); $t(23)=.912;$ $p=.371;$ $95\%$ $CI[-8.90, 22.95]$) nor the tunnel condition ($task_{energy}=51.35$ (SD=14.98); $task_{garbage}=57.15$ (SD=12.85); $t(23)=-1.034;$ $p=.312;$ $95\%$ $CI[-17.39, 5.8]$). Even though means differ by 10\%, we attribute this result to chance, as the sign differs between both locomotion techniques. This observation speaks against a structural difference between both tasks. Thus, we treat both tasks as equivalent for further analysis.

Consequently, we compared only the evaluated measures between both within-conditions tunnel and teleport with paired sample t-tests. Beforehand, we ensured the test's assumptions by testing for normal distribution with Shapiro-Wilk tests. All listed calculations were executed with IBM SPSS 27~\cite{spss}. In the following section, we report the significant differences between conditions, including all necessary information, such as the effect strength and the confidence interval, to ensure reproducibility~\cite{vornhagen2020statistical}.

\subsection{Questionnaires}

In order to confirm our hypothesis H1, we assessed the SSQ. The resulting weighted scores are shown in Table~\ref{tab:results}. They indicate no noteworthy cybersickness symptoms according to reference values by Kennedy et al.~\cite{Kennedy.1993} Besides, both conditions do not differ significantly, and the small confidence intervals of the means suspend the presence of a larger undiscovered effect. Furthermore, we measured the perceived presence between both conditions to answer RQ2. Whereas most subscales of the PQ (see Table~\ref{tab:results}) are almost equal for the teleport and our approach, the paired sample t-tests for the total presence score and the realism subscale indicate an observed medium effect according to Cohen's d. However, as the confidence intervals include both trivial and meaningful differences, we cannot reach a decisive conclusion concerning the effect's nature. Finally, we assessed the PXI subscales mastery, autonomy, and ease of control. Only the statistically significant differences for ease of control suggest the presence of a  medium effect.

\begin{table*}[t]
  \caption{Mean scores, standard deviations, and paired sample t-test values of the custom questions (CQ).}
  \label{tab:custom}
  \begin{tabularx}{\textwidth}
  {>{\raggedright\arraybackslash}p{0.6cm}
  >{\raggedright\arraybackslash}p{4.4cm}
  >{\centering\arraybackslash}X
  >{\centering\arraybackslash}X
  >{\centering\arraybackslash}p{1.2cm}
  >{\centering\arraybackslash}p{0.55cm}
  >{\raggedright\arraybackslash}p{0.15cm}
  >{\centering\arraybackslash}p{1cm}
  >{\centering\arraybackslash}p{2.3cm}
  }
    \toprule
     && Tunnel & Teleport \\
    \multicolumn{2}{l}{Question Item} & M(SD) & M(SD) &$t(24)$ & $p$ & & $d$ & CI\\
    \midrule
    CQ1 & I could orient myself well in the virtual world.  & 4.12 (1.51) & 4.68 (0.95) & -2.064 & .050 & & -.413 & $[-1.120,.000]$\\
    CQ2 & After each relocation, I needed a moment to orient myself. & 3.12 (1.81) & 2.56 (1.85) & 1.184 & .248 & & .237 & $[-.416,1.536]$\\
    CQ3 & I gained a good impression of the traveled distance. & 2.96 (1.62) & 3.12 (1.39) & -.458 & .651 & & -.092 & $[-.882,.562]$\\
    
    CQ4 & I felt very active while playing.  & 4.08 (1.41) & 3.32 (1.57) & 2.282 & .032 & * & .456 & $[.073,1.447]$\\
    CQ5 & I think I have been walking much in the real room while playing. & 4.48 (1.45) & 3.28 (1.54) & 2.969 & .007 & ** & .594 & $[.366,2.034]$\\
    
    CQ6 & Traveling longer distances was cumbersome. & 0.60 (0.87) & 1.36 (1.29) & -2.317 & .029 & * & -.463 & $[-1.437,-.083]$\\
    CQ7 & I would have preferred to move through the world using another technique. & 2.24 (1.64) & 2.56 (1.56) & -.858 & .399 & & -.172 & $[-1.089,.450]$\\
    CQ8 & I was able to control the navigation between the different locations. & 1.96 (1.54) & 4.12 (1.39) & -5.014 & .000 & ** & -1.003 & $[-3.049,-1.271]$\\
    \bottomrule
     &&&&&&  \multicolumn{3}{r}{*\textit{p} $<.05$, ** \textit{p} $<.01$}\\
\end{tabularx}
\end{table*}

\subsection{Custom Questions}

Besides the aforementioned standardized questionnaires, we also administered eight custom questions to confirm our hypothesis H3 and gain further insights into the personal user experience. These questions were split into three parts: The first three questions covered spatial orientation. Next, we included three questions on physical activity and closed with two general questions on personal preference and autonomy. The results for all questions are shown in Table~\ref{tab:custom}. The items concerning orientation are not statistically significant. In contrast, all three questions covering the feeling of activity imply the presence of medium effects. It appears that subjects felt more active when using the tunnel approach without experiencing it as cumbersome. Finally, the last question reveals that the subjects felt far less autonomous when using our presented concept.

\subsection{Logging Data and Distance Estimations}

\begin{figure*}[t!]
\centering
\includegraphics[width=0.8\columnwidth]{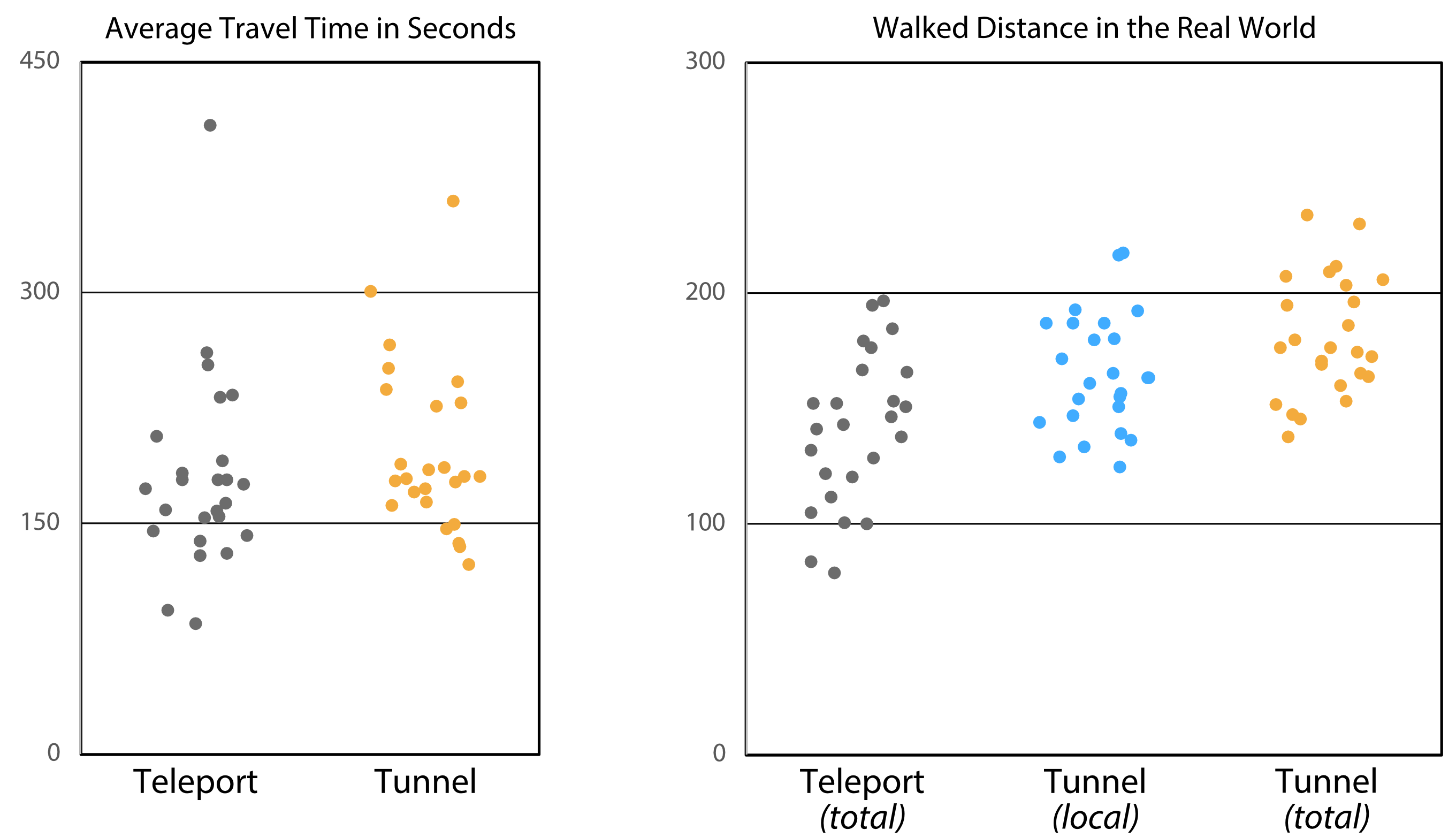}
\caption{Results from the data logged during the study. Left: the difference in travel time for both conditions in seconds. Right: the average distance (in meters) subjects walked in the real room. For the tunnel condition, \textit{local} indicates the distance subjects walked at the local checkpoints. In contrast, \textit{total} measures the sum of local navigation and the distance subjects walked through the tunnels. For the teleport condition, only the walked distance in \textit{total} is provided.}
\label{fig:diagram}
  \Description{Two scatterplots visualizing the logged data. The left scatterplot, indicating the average travel time, reveals widespread but no major differences between the conditions. All three data categories in the right scatterplot have similar spreads. However, the walking distance is generally the lowest for the teleport, about $~10\%$ higher for the local navigation of the tunnel concept, and another $~10\%$ higher when adding the tunnel travel distance.}
\end{figure*}

For H2, we assumed that our augmented-walking approach provokes the subjects to walk more while traveling longer. To confirm this hypothesis, we logged the necessary data throughout both rounds (see Figure~\ref{fig:diagram}). Whereas the average total travel time between checkpoints for the tunnel condition was 9.33\% higher, this difference was not statistically significant ($t(24) = 1.236;$ $p = .228;$ $d = .247;$ $95\%$ $CI[-11.255,44.865]$). However, subjects using the tunnel concept walked 28.37\% more compared to the teleport condition. This difference is highly significant and implies a very large effect ($t(24) = 6.292;$ $p < .001;$ $d = 1.258;$ $95\%$ $CI[26.850,53.065]$). Even after subtracting the necessary walking distance through the tunnels, the difference in the remaining local walking distance remains significant ($t(24) = 3.916;$ $p < .001;$ $d = .783;$ $95\%$ $CI[11.584,37.396]$). This finding reveals that subjects also tended to rely on teleportation overly and thus walked 14.81\% less while completing the local tasks.

Finally, the subjects also had to estimate their traveled distance from one checkpoint to another. We aggregated these individual estimations by calculating the mean squared error (MSE) compared to the correct distance. In this step, we had to exclude three extreme outliers with MSEs above 1300. The remaining data suggest that the teleport technique leads to slightly better distance estimations: $MSE_{teleport}=322.64$ (SD=204.49), $MSE_{tunnel}=406.75$ (SD=245.89). However, this difference is not statistically significant ($t(21) =- 1.641;$ $p =.116;$ $d = -.350;$ $95\%$ $CI[-190.735,22.508]$). Examing the unsquared errors (ME) reveals that the subjects generally underestimated the distance: $ME_{teleport}=-2.45$ (SD=13.56), $ME_{tunnel}=-6.14$ (SD=13.52).

\section{Discussion}

In general, all subjects enjoyed participating in our study and could complete all tasks without difficulties. Whereas most subjects were familiar with the teleport concept, none had previously used an accelerated walking approach. Nevertheless, multiple subjects reported past experiences with cybersickness when using artificial navigation concepts, such as gamepad locomotion. As such symptoms cause unwellness that can retain for hours, some subjects were extra cautious when trying VR applications. Consequently, the avoidance of cybersickness was our main priority, followed by two further hypotheses and research questions covering the relevant qualities of locomotion techniques.

\subsection*{H1: The presented augmented-walking approach does not induce cybersickness.}

According to the results of the SSQ, our locomotion technique did not cause cybersickness symptoms. Moreover, the observed values are equal to the ones of the teleport condition. This finding is promising as teleportation is usually also chosen for its superior tolerability for sensitive users. Furthermore, we did not observe any adverse effects despite applying a large gain factor to the virtual movement, which induces high visual flows and causes cybersickness. This finding is to be emphasized as prior research reported the occurrence of cybersickness symptoms already at gain factors below $10x$~\cite{tirado2019_gains}. In sum, these results confirm our primary hypothesis H1.

Despite these promising results, there was a single case of mild symptoms of dizziness after using the tunnel technique. Our subsequent interview suggests that the observed indications might result from the subject testing the limitations of our navigation concept. For instance, the respective subject repeatedly leaned forward and back in the tunnel's center while looking through the windows. Even though this behavior was not intended, we emphasize the need for further investigation. A possible solution to this drawback might be to use center-of-mass displacement instead of head displacement for the movement scaling. Finally, we only assessed the SSQ after the study condition and not additionally in advance. Whereas this study design is used in a variety of prior studies~\cite{freitag2014_portals, laviola2000_cybersickness, krekhov2018_gullivr}, it might have introduced a confounding factor, as we cannot account for the pre-study state of the subjects.

\subsection*{H2: Compared to teleportation, our locomotion technique significantly increases the walked distance and the overall travel time.}

For the tunnel condition, we used our presented walking-based travel approach paired with real walking for the local tasks. This combination increased the total walked distance significantly. What is more, it also permitted us to avoid using virtual locomotion techniques in this condition. As a consequence, the subjects also walked significantly more while completing the local tasks. In contrast to this outcome, the teleport concept encouraged the subjects to rely on teleportation exclusively, regardless of whether a target was reachable by real walking or not. This finding was already reported in other papers~\cite{cmentowski2019_outstanding, krekhov2018_gullivr, freitag2014_portals} and is also reflected in the custom questions as the subjects felt more active in the tunnel condition. 

We also assumed that the observed extended walking would increase the total travel time compared to the instant teleport. However, we could not confirm this part of the hypothesis. Instead, both concepts required a similar time to travel between the checkpoints. It seems that the tunnel locomotion, i.e., waiting for the tunnel animation before walking $1.5m - 2.5m$, does not take longer than aiming and teleporting multiple times.

\subsection*{H3: The introduced tunnel concept significantly benefits spatial orientation and gives a better impression of the traveled route compared to teleportation.}

We assessed the difference in spatial orientation and overview using three custom questions and the distance estimations. However, these measures could not confirm our hypothesis. Instead, the means of the questions and MSEs point slightly towards the teleport for providing better orientation. Nevertheless, this difference is not significant. We assume that the reasons for this equality mainly reside in the tunnel's geometry. Although being necessary to prevent cybersickness, the tunnel's walls hide most of the surroundings during travel. We added the windows to diminish the adverse effect of the partial occlusion. However, they might not have sufficed to deliver the benefits of real walking. Still, the performance of our novel locomotion concept is comparable to the teleport, which is one of the most commonly used techniques. In future research, we plan to investigate this topic further by finetuning the window slits for an optimal view and conducting a study with a dedicated orientational task.

\subsection*{RQ1: How does our locomotion technique compare to teleportation regarding mastery and ease of control?}

We did not formulate a confirmatory hypothesis for the locomotion technique's usability, since we compare our approach against teleport, which is known for its simplicity and easy usage~\cite{bozgeyikli2016_teleport}. Still, the tunnel concept does not require prior training by relying solely on real walking. To investigate how both techniques compare, we assessed three PXI subscales and two custom questions. Both techniques were rated equally popular (CQ7) and easy to master. However, the tunnel concept appears to be slightly easier to control than the teleport. This significant result of the PXI reflects verbal feedback of multiple subjects, one of whom stated, \textit{"it took time to get used to the two controller buttons for object manipulation and teleporting"(\textbf{P10})}. At the start of the teleport condition, few subjects confounded the inputs and accidentally teleported themselves. Thus, the tunnel might be easier for beginners by employing known concepts.

\subsection*{RQ2: Do both navigation approaches differ concerning perceived presence?}

Finally, we were interested in how the locomotion techniques influence the perceived presence. In general, most PQ subscales are almost identical. The two notable differences are the realism subscale and the total presence, which are both significantly higher for the tunnel condition. Of course, the presence sensation is very individual and highly scenario-dependant. However, some subjects stated that \textit{"the teleport felt a bit too unrealistic by happening instantaneously"(\textbf{P16})}. The tunnel had no such visual cut and felt smoother in movement. 

\section{Design Considerations and Limitations}

Our proposed navigation concept extends on existing augmented-walking approaches and effectively prevents cybersickness by hiding the increased visual flow. Thereby, it adds to the growing collection of existing locomotion techniques for virtual environments. In this section, we discuss the decisive design considerations and limitations concerning the application of our approach for VR experiences.

\subsubsection*{Benefits of a Walking-based Locomotion Technique}

As explained earlier, using real walking for VR locomotion causes the users to feel more active~\cite{cmentowski2021_sneaking} and potentially boosts spatial orientation~\cite{ruddle2011_walking}. Two additional benefits were mentioned by our study subjects. Multiple subjects reported an increased feeling of safety while using the tunnel concept. When using teleportation, \textit{"one would often end up in one corner of the real room, without enough space to interact with the environment"(\textbf{P4})}. In contrast, the tunnel condition used the real playspace effectively so that \textit{"one could use all tunnels and solve every task without even seeing the collision warning once"(\textbf{P1})}. This benefit is achieved in this particular case by designing the virtual environment with the real surroundings in mind, which requires knowledge of the playspace in advance and is also impossible for free teleport locomotion. Finally, one subject stated another simple reason for preferring the walking-based concept: \textit{"Without the teleport, I could use both hands at once and carry two items at the same time"(\textbf{P2})}.

\subsubsection*{Choosing the Gain Factor}

For our study, we selected a fixed translational gain factor of $30x$. This decision was based on the size of the virtual world and the available tracking space. Thus, other applications likely require different gain factors to use our locomotion technique effectively. Even though we cannot generalize our results to the performance of other translational gains, our concept still avoids most of the issues observed in previous research with values lower than $10x$. In general, two approaches for realizing different gains are feasible: In our case, we preselected a fixed gain factor, which produced tunnels of various lengths, ranging from $1.5m$ ($45m$ virtual distance) to $2.5m$ ($75m$). Alternatively, one could use a standardized tunnel distance, such as $2m$, and vary the gain factor based on the virtual path. The final option is to adapt the gain factor according to the remaining physical space in front of the user to use the playspace more effectively. Extending this approach to curved tunnels could be used to steer players away from the walls and thereby minimize potential resets. Regardless of this design decision, future work is still needed to identify possible lower and upper bounds to the gain factor concerning cybersickness and travel experience.

\subsubsection*{Setting the Navigation Target}

Opposing the point and teleport concept, our locomotion technique works along a straight path from the users' current position to a predefined destination. However, this design requires setting the target position before deploying the tunnel. In our study, we used the tubular level design to set fixed navigation goals between the checkpoints. This design decision limits the players' freedom to navigate freely, which is clearly reflected in the results of CQ8. However, this experience did not negatively prejudice the overall game experience, as seen in the equal scores of the PXI's autonomy subscale. Past research emphasized that the primary use case of large translational gains mainly resides within "large displacements along non-relevant areas"~\cite{montano2019computational}. Many games employ a similar environmental design of distinct points of interest and neglectable transitional areas, and we argue that our concept is particularly suited for such scenarios. However, freely explorable environments require additional user interaction to set the next locomotion target. Visual pointers, used for various navigation techniques, perform poorly for long distances due to the so-called "fishing-rod problem". Instead, dedicated visualizations, such as minimaps or worlds-in-miniature, provide a better solution, preserving accuracy regardless of the travel distance. These concepts are also compatible with fixed navigation anchors to combine self-determined locomotion with the benefits of a pre-planned walking experience.

\subsubsection*{Restrictions on the Virtual Environment}

By design, our augmented-walking concept does not require strict design characteristics with respect to the surrounding virtual world. It works well in a variety of environments, ranging from broad and open scenarios to narrow indoor areas. The only decisive prerequisite is the straight, occlusion-free path between start and end position. In future work, we aim to investigate a possible extension to curved routes. In this case, the tunnel's cabin would remain rectangular but follow the path's course during locomotion. Also, we did not investigate whether the direct surrounding influences the users' travel experience. Narrow paths might provide a poorer impression of the surroundings than broad areas. Although we eliminated this factor in our study by using similar level geometry, future exploration of possible influences is decisive for broader application.

\subsubsection*{Restrictions on the Playspace}

Apart from limitations to the surrounding virtual scenario, our tunnel concept requires at least a medium-sized play area. Generally speaking, larger spaces allow for longer tunnels without reaching the physical boundaries at every tunnel entry and exit. The necessity to entirely leave the tunnel before it disappears significantly reduces the maximal tunnel length compared to the available space. For instance, for our $16m^2$ lab, we decided on a maximal tunnel length of $2.5m$ to avoid complicated maneuvering in the playspace's corners. To a certain degree, short tunnels can be compensated for by increasing the gain factor further. However, this design decision risks diminishing the players' impression of the virtual travel and should be used with caution. In the end, the usability and effectiveness of the tunnel technique depend on the ratio between virtual and real travel distances. Consequently, we recommend this locomotion technique only for medium to large playspaces, starting at about $16m^2$.

\subsubsection*{Windows}

Apart from preventing cybersickness, our novel tunnel concept also aimed to provide an optimal travel experience through windows in the tunnel's walls. Our chosen design --- horizontal stripes --- was carefully chosen in a participatory design phase to combine a good view with a steady visual anchor point. Whereas the concept generally worked well, the spatial orientation was similar to the teleport condition and not superior as expected. One potential reason might reside in the age difference between the participants of our participatory design phase and the subjects of our study. As we optimized the size and shape of the window slits according to the feedback of our early testers, these properties might not have been optimal for the larger player base in the main study. Additionally, we only measured differences in spatial cognition through questionnaires. For an advanced understanding of the underlying effects, alternative measurements are preferable, such as pointing tasks or map-drawing tasks. Therefore, future research should approach these shortcomings to determine the optimal window size that combines cybersickness-free augmented walking with optimal spatial knowledge.

\subsubsection*{Embedding and Animations}
We used a thematic style and transitional animations for our implementation and user study. These optional design elements are primarily cosmetic and have a minimal effect on the actual locomotion experience. Nevertheless, they offer valuable advantages. Firstly, Marwecki et al.~\cite{marwecki2018virtualspace} argue that techniques, such as the Seven-League Boots concept, involve an overt deviation between physical movements and virtual locomotion, which could reduce "the immersive quality of real walking". In contrast, our movement acceleration occurs within the tunnel's static geometry and ends when reaching the exit. This diegetic embedding transforms the dedicated locomotion technique into a feature of the virtual environment, leaving the original one-to-one mapping of real walking intact.

Furthermore, previous augmented walking concepts were only experienceable after the users began walking and could easily lead to confusion. Instead, our tunnel's geometry and the animated doors emphasize a clear cut between the locomotion modes. However, invoking the tunnel and waiting for the intro animation pose a considerable interaction overhead. In our study, no participant was bothered by this design, and both locomotion techniques also performed similarly despite this overhead. Nevertheless, repeated use of our technique in longer play sessions might annoy players. Thus, we propose adapting the animation speed in subsequent invocations after an initial adaptation phase.

\section{Conclusion}

Tracking the users' movements in the real playspace and translating them into the virtual world is widely considered the gold standard for VR locomotion. However, the physical constraints of the available tracking space usually limit the applicability of physical walking for larger environments. One possible solution is to scale these movements and thereby enable users to walk further in VR. However, such augmented virtual motions increase the visual flow and are known to induce cybersickness. Also, established concepts often suffer from poor accuracy and implementation hurdles. Our presented technique is a novel alternative that uses a tunnel concept to shield the players from excessive visual flow. While the players walk through this seemingly short tunnel, they perceive only their accelerated forward motion through windows in the tunnel's walls.

In contrast to the well-known teleport technique, our approach increases the perceived and actual physical activity while effectively preventing cybersickness and preserving high levels of presence. Also, our within-subject study revealed that the concept is more beginner-friendly than the teleport approach. In the last part of the paper, we summarized the resulting design considerations and the limitations of our technique. Our future research will focus on the various open questions and possible improvements explained in the previous section, including a closer investigation of different gain factors and window shapes and the exploration of curved paths and self-set navigation targets.

\bibliographystyle{ACM-Reference-Format}
\bibliography{literature}

\received{February 2022}
\received[revised]{June 2022}
\received[accepted]{July 2022}

\end{document}